\documentstyle[12pt,epsfig,aps,prc]{revtex}
\begin{document}

\title{Complete determination of the reflection coefficient in neutron
specular reflection by absorptive non-magnetic media
}
\author{H. Leeb, H. Gr\"otz, J. Kasper}
\address{Institut f\"ur Kernphysik, Technische Universit\"at Wien,
Wiedner Hauptstra\ss e 8-10/142, A-1040 Vienna, Austria}
\author{R. Lipperheide}
\address{Hahn-Meitner-Institut Berlin und Freie Universit\"at Berlin,
Glienicker Strasse 100, D-14091 Berlin, Germany} 
\date{\today}
\maketitle

\begin{abstract}
An experimental method is proposed which allows the complete 
determination of the complex reflection coefficient for absorptive 
media for positive and negative values of the momenta. It makes use 
of magnetic reference layers and is a modification of a recently 
proposed technique for phase determination based on polarization 
measurements. The complex reflection coefficient resulting from a
simulated application of the method is used for a reconstruction of 
the scattering density profiles of absorptive non-magnetic media by 
inversion.
\end{abstract}

%
%
%
%
\section{Introduction}

Neutron specular reflection has become a common tool in materials sciences
\cite{Felcher91}. Nevertheless, there are still severe difficulties in the 
interpretation of reflectometer experiments because of the so-called 
{\em phase problem}. This problem refers to the incompleteness of the data
obtained in standard
experiments in structure physics \cite{Buerger67,Hauptmann91} where only the 
intensities of the reflected waves are measured but not the corresponding 
phases. In specular reflection both the modulus and the phase of the reflection
coefficient are needed for an unambiguous reconstruction of the surface 
profiles \cite{Klibanov92,Reiss96}. 

Several solutions of this so-called {\em phase problem} have been proposed 
for neutron specular reflection 
\cite{Klibanov92,Fiedeldey91,Majkrzak95,Haan95,Leeb98,Kasper98}.
The recently proposed reference layer methods based on polarization
measurements \cite{Leeb98,Kasper98} are of particular interest because
they also work in the total reflection regime and allow the unique
reconstruction of surface profiles of magnetic samples. Despite the
great effort in theory and modelling, only the phase determination by 
the method of Majkrzak and Berk \cite{Majkrzak95} was tested experimentally
 \cite{Majkrzak98} for a specific case.

The available methods of phase determination
[e.g. \cite{Majkrzak95,Haan95,Majkrzak98,Majkrzak98b}] for non-absorptive 
as well as for absorptive samples provide us with the full (modulus and phase) 
reflection coefficient $R(q)$ for positive values of the wave number $q$ 
perpendicular to the surface of the sample. On the other hand, the 
reconstruction of the surface profiles by inversion requires the
knowledge of $R(q)$ over the positive {\it and} negative range of 
$q$-values. This is no problem for non-absorptive samples 
because in this case the reflection coefficient $R(q)$ satisfies
\begin{equation}
R(-q)=R^\dagger (q)
\, ,
\label{unitary}
\end{equation}
and $R(q)$ at negative $q$-values is directly given in terms of 
$R(q)$ at positive $q$. However, in the presence of absorption relation 
(\ref{unitary}) does not hold, and therefore a reconstruction of 
the surface profiles is not possible without further input. To our 
knowledge, so far no procedure has been proposed for 
providing the missing information.

In this work we propose a modification of the reference layer method
for non-magnetic samples based on polarization measurements \cite{Leeb98}.
The modification consists in the interchange of the positions of the
sample and the reference layer, which allows the determination of
the reflection coefficient for positive and negative $q$-values. 
The method works also in the total reflection regime (in contrast to
\cite{Majkrzak95,Haan95,Majkrzak98}); it is only limited 
by the sample thickness because of the corresponding Kiessig oscillations 
which must be resolved. 

In section II we present the method and derive the basic relations 
for the phase determination. A realistic example is given in section III, 
where we demonstrate by simulations the feasibility of the method. The 
problem of thick samples is also discussed. A brief summary and concluding 
remarks are given in section IV.

\section{The method}

We consider the arrangement of an unknown non-magnetic sample and a magnetic 
reference layer mounted on a substrate (e.g. a Si-wafer) as shown in Fig. 1. 
We assume that within the reference layer ($0\leq x\leq a$) there is a 
magnetic field ${\bf B}(x)$ aligned with the surface of the sample. The 
direction of this field is taken as the $z$-axis, which is chosen as the 
axis of spin quantization. 
The direction of propagation perpendicular to the surface defines the 
$x$-axis in a right-handed coordinate system. Such a magnetic 
reference layer may consist of a ferromagnetic stratum, e.g. a Fe-, 
Co- or Ni-layer (cf. Fig. 1). We assume that the sample as well as 
the substrate are field-free. However, the further considerations remain
still valid as long as the condition $|{\bf B}(x)| = 0$ for $x \to \infty $ 
is satisfied. 


The arrangement of Fig. 1 differs from that of \cite{Leeb98} only by the
interchange of sample and reference layer and therefore we can make use 
of the same relationships for the description of the reflection. 
The position of the sample on top of the reference layer and the substrate, 
however, leads to a completely new situation as compared to previous 
proposals for the solution of the phase problem. This is easily seen from 
the expression for the reflection coefficient $R^{tot}$  of the whole 
arrangement  
\begin{equation}
R^{tot}_L(q) = \frac{E(q)\rho_L(q) + R_L(q)}{1 - R_R(q) \rho_L(q)} \, .
\label{eq:super}
\end{equation}
Here the reflection properties of the sample enter in terms of the 
reflection coefficients $R_L$ and $R_R$ and the quantity
\begin{equation}
E(q)= T_L(q)T_R(q) - R_L(q) R_R(q) \, ,
\label{eq:E}
\end{equation}
where $T_L, T_R$ are the transmission coefficients of the sample. The
quantity $\rho_L$ is the reflection coefficient of the reference 
layer plus the substrate and the indices $L, R$ refer to incidence  
from left and right, respectively. Because of the rather
intricate dependence of $R_L^{tot}$ on the reflection properties of
the sample the separation of its unknown reflection coefficient becomes 
more involved.

Following the procedure of \cite{Leeb98} we introduce the quantity
\begin{equation}
s = \frac{R_{L+}^{tot}}{R_{L-}^{tot}}\, ,
\label{eq:sdef}
\end{equation}
 which can be expressed in the form
\begin{equation}
s = \frac{P_x^0+iP_y^0}{P_x+iP_y}
\frac{1+P_z}{1+P_z^0}\, ,
\label{eq:s}
\end{equation}
where ${\bf P}=(P_x,P_y,P_z)$ and ${\bf P}^0=(P_x^0,P_y^0,P_z^0)$ are 
the polarizations of the reflected and the incident beam, respectively.
The indices $\pm $ refer to neutron beams polarized parallel $(+)$ and
antiparallel $(-)$ to the magnetic field. All quantities on the 
right-hand side of Eq. (\ref{eq:s}) are measurable and therefore $s$ can 
be determined experimentally.

Use of Eq. (\ref{eq:super}) for $R_\pm^{tot}$ in Eq. (\ref{eq:sdef}) leads 
after simple algebraics to the relation
\begin{equation}
- \frac{R_L(q)}{E(q)} = 
\frac{(\rho_L^+ - s\rho_L^-)-R_R\rho_L^+\rho_L^-(1-s)}
{(1-s)-R_R (\rho_L^- - s \rho_L^+)}
\, .
\label{eq:Rn}
\end{equation}
The term on the left-hand side of Eq. (\ref{eq:Rn}) is equal to $R_R(-q)$,
i.e. to the right reflection coefficient at negative $q$-values 
(cf. \cite{Lipperheide95,Lipperheide96}). This is easily 
seen from the Jost solutions of the bare sample. Making use of the linear 
independence of the Jost solutions $f_{L,R}(q,x)$ with asymptotic forms
\begin{equation}
f_L(q,x)=\left\lbrace \begin{array}{l} 
\left[ e^{iqx}+R_L(q) e^{-iqx}\right]/T_L(q) \\ e^{iqx} \end{array} 
\right. \begin{array}{l} 
\mbox{for } x \to -\infty \\\mbox{for } x \to +\infty \end{array}
\label{eq:JostL}
\end{equation}
and
\begin{equation}
f_R(q,x)=\left\lbrace \begin{array}{l}
e^{-iqx}\\ \left[ e^{-iqx}+R_R(q) e^{iqx}\right]/T_R(q) \end{array}
\right. \begin{array}{l}
\mbox{for } x \to -\infty \\ \mbox{for } x \to +\infty \end{array}
\, ,
\label{eq:JostR}
\end{equation}
one finds
\begin{equation}
f_R(-q,x)=T_L(q)f_L(q,x)-R_L(q)f_R(q,x) \, .
\label{eq:fR}
\end{equation}
Comparing Eq. (\ref{eq:fR}) in the limit $x \to \infty $ with the 
corresponding asymptotic form (\ref{eq:JostR}) with $q$ replaced by $-q$
we obtain 
\begin{equation}
R_R(-q)=- \frac{R_L(q)}{E(q)}
\, ,
\end{equation}
and therefore, from Eq. (\ref{eq:Rn}).
\begin{equation}
R_R(-q) = 
\frac{(\rho_L^+ - s\rho_L^-)-R_R(q)\rho_L^+\rho_L^-(1-s)}
{(1-s)-R_R(q) (\rho_L^- - s \rho_L^+)}
\, .
\label{eq:Rmn}
\end{equation}
This is a relation involving measurable quantities which expresses
the right reflection coefficient at $-q$ ($q > 0$) in terms of its
value at $q$.

For a full determination of the reflection coefficient $R_R(q)$ at 
positive and negative $q$-values (i.e. for obtaining two independent
equations for the two unknowns $R_R(q)$ and $R_R(-q)$) at least two
sets of measurements with different reference layers are necessary.
Denoting the two measurements with the upper indices $^{(a)}$ 
and $^{(b)}$ one obtains from Eq. (\ref{eq:Rmn}) the quadratic 
equation for $R_R(q)$
\begin{equation}
\alpha +\beta R_R(q) + \gamma R_R^2(q) = 0
\label{eq:quadr}
\end{equation}
with
\begin{equation}
\alpha = 
(1-s^{(b)})(\rho_L^{+ (a)}-s^{(a)} \rho_L^{- (a)})-
(1-s^{(a)})(\rho_L^{+ (b)}-s^{(b)} \rho_L^{- (b)})
\, ,
\label{eq:alpha}
\end{equation}
\begin{eqnarray}
\beta & = &
(\rho_L^{- (b)}-s^{(b)}\rho_L^{+ (b)}) 
(\rho_L^{+ (a)}-s^{(a)}\rho_L^{- (a)}) -
(\rho_L^{+ (b)}-s^{(b)}\rho_L^{- (b)})
(\rho_L^{- (a)}-s^{(a)}\rho_L^{+ (a)})
\nonumber \\
& + &
(1-s^{(a)})(1-s^{(b)})\rho_L^{+ (a)}\rho_L^{- (a)}-
(1-s^{(a)})(1-s^{(b)})\rho_L^{+ (b)}\rho_L^{- (b)}
\, ,
\label{eq:beta}
\end{eqnarray}
\begin{equation}
\gamma = 
(1-s^{(a)})(\rho_L^{- (b)}-s^{(b)} \rho_L^{+ (b)})\rho_L^{+ (a)}
\rho_L^{- (a)} -
(1-s^{(b)})(\rho_L^{- (a)}-s^{(a)} \rho_L^{+ (a)})\rho_L^{+ (b)}
\rho_L^{- (b)}
\, .
\label{eq:gamma}
\end{equation}
The quadratic equation (\ref{eq:quadr}) has two roots
\begin{equation}
R_R^{(1,2)}(q) = \frac{-\beta \pm \sqrt{\beta^2-4\alpha \gamma}}{2\gamma }
\, .
\label{eq:root}
\end{equation}
Similarly to the procedure of Ref. \cite{Leeb98} the physical solution
$R_R(q)$ can be selected either by continuity requiring that the phase 
$\Phi (q)$ satisfies $\phi (q=0)=-\pi $ or by the condition 
$r_R=|R_R|^2\leq 1$. The latter may fail at some $q$-values and must
be accompanied by continuity in certain momentum regions. The selection 
by continuity implies measurements of $s$ over the whole range of 
momenta. The solution $R_R(-q)$ at negative $q$-values is given by
Eq. (\ref{eq:Rmn}).

We note that the method relies solely on measurements of the 
polarization of the reflected beam. In particular, even for absorptive
samples, no transmission measurements are needed for the determination of 
$R(-q)$ as might be conjectured from the general formalism 
\cite{Lipperheide96}.

\section{Example}

We test the method by a simulation using the realistic example shown in 
Fig. 1. The sample consists of a $30$\ nm Au- and a $10$\ nm Cd-layer, 
where the latter is strongly absorptive. To determine the reflection 
coefficient for positive and negative $q$-values we consider measurements 
with two different reference layers. Both reference
layers are composed of a $15$\ nm Cr-layer and a $15$\ nm thick film
of ferromagnetic material magnetizied up to saturation and mounted on a 
Si-wafer. The required difference in the reflection properties of the
reference layers can be achieved by using Fe in measurement (a) and 
Co in measurement (b) for the ferromagnetic layer. The magnetization
of this layer will generate a magnetic induction also outside the
ferromagnetic film which we assume, however, to be small enough not to 
affect the neutron beam. For simplicity we set it equal to zero.

The reflectivity $r_{L+}^{tot}=|R_{L+}^{tot}|^2$ and the polarization
components $P_x, P_y, P_z$ for the arrangement of Fig. 1 have been 
calculated for both reference layers at all positive $q$-values. To
illustrate the typical observables which have to be measured for the 
phase determination we show in Fig. 2 the reflectivity 
$r_{L +}^{tot}=|R_{L +}^{tot}|^2$ and the components of the polarization 
$P_x, P_y, P_z$ for the reference layer (a) assuming a fully polarized 
incident beam in $x$-direction. It is evident that in the given example
measurements in the range $0\leq q\leq 1.0$\ nm$^{-1}$ are certainly
feasible. 
%

Following the procedure outlined in section II we obtain two solutions
for $R_R(q)$ which are displayed in Fig. 3. As expected, in a wide range 
of positive $q$-values only one solution satisfies $|R_R(q)|\leq 1$ and 
is therefore physically admissible. Applying continuity the region of 
uniqueness can be extended over all positive $q$-values. In addition
one obtains via Eq. (\ref{eq:Rmn}) unique values of $R_R(-q)$. Hence 
the measurements of the polarization at positive $q$-values is sufficient
for the determination of the complete  reflection coefficient on the
whole $q$-axis.
%

To extract the surface profiles from the extracted reflection coefficients
$R_R(q)$ we apply the Marchenko inversion procedure used in Ref.
\cite{Lipperheide95}, where it was shown to yield correct reconstructions
of the {\em complex} input potential, although a rigorous mathematical 
proof of this is lacking. The Marchenko inversion procedure is usually 
formulated for $R_L(q)$, while our procedure provides $R_R(q)$. 
Because of the specific choice of our coordinate system we can also use 
$R_R(q)$ in the formulation of Ref. \cite{Lipperheide95} but we have to 
change the argument of the evaluated potential from $x$ to $-x$. This 
procedure has been applied to analyse the numerically extracted values of 
$R_R(q)$. The results for the scattering length density profile are 
displayed in Fig. 4. It is obvious that the method reproduces the original 
potential within the resolution determined by the maximum available 
$q$-value. 

Since the method is strongly related to the reference layer method presented
previously \cite{Leeb98} its stability with respect to experimental 
uncertainties, e.g. roughness of interfaces and measurement errors, is
similar to that of method \cite{Leeb98}. This is also true for the 
reconstruction of the surface profiles. For more details on the stability 
we refer to our previous studies (cf. \cite{Leeb98,Lipperheide98}).

\section{Conclusions}

We have proposed a method involving a magnetic reference layer which,
for the first time, allows the determination of the full complex 
reflection coefficient at positive and negative $q$-values. Thus a 
reconstruction of surface profiles of absorptive non-magnetic samples 
becomes feasible. The novel procedure is a modification of the previously 
proposed method \cite{Leeb98}. The main difference is the interchange of 
the position of the sample and the reference layer. Because of this change 
the reflection properties of the sample enter in a more involved way and 
require two sets of measurements using two different reference layers. 
The additional effort is remunerated by providing us with an additional 
relationship from which the reflection coefficient at negative $q$-values
can be obtained. This novel feature allows the determination of
absorptive surface profiles by inversion, -- an open problem up to now.
In a numerical example it is demonstrated that the reconstruction of real and 
imaginary parts of the scattering density profiles is feasible. 

As in the previously proposed reference layer methods 
\cite{Majkrzak95,Haan95,Leeb98} the thickness of the layers is
reflected in the $q$-dependence of the measured data. Specifically,
the so-called Kiessig oscillations must be resolved in experiment.
Hence, the degree of the monochromaticity as well as the resolution
of the diffraction angle of the incident beam set a limit on the
thickness of the studied system. The roughness of the interfaces has
similar effects and results in additional uncertainties in the deduced 
reflection coefficient (cf. \cite{Leeb98}).

In summary the phase problem of neutron reflection has been solved for
absorptive non-magnetic samples. The method seems experimentally feasible
for sample thicknesses up to $100$\ nm and complements the reference 
layer methods \cite{Leeb98,Kasper98} recently presented.


%
%
%
%
%
%
\vfill
\newpage

\section*{Figure Captions}

{\bf Figure 1}\\
Experimental arrangement for measuring the complex reflection coefficient
Top: Arrangement of the layers. Bottom: The potential profile; the real
part is represented by the solid lines, the imaginary part by the dashed 
area. The dotted lines represent the effective potentials experienced by
neutron beams polarized parallel and antiparallel to the magnetic field 
${\bf B}$.

\vspace{0.5cm}
{\bf Figure 2}\\
Simulated reflectivity and polarization data for the arrangement of Fig. 1 
with a magnetized Fe-film in the reference layer. The incident beam is 
assumed to be fully polarized in the $x$-direction. The reflectivity 
$r_{L+}^{tot}=|R_{L+}^{tot}|^2$ (a) and the polarization components  
$P_x$ (b), $P_y$ (c) and $P_z$ (d) of the reflected beam are shown.

\vspace{0.5cm}
{\bf Figure 3}\\
The roots $R_R^{(1,2)}$ of the quadratic equation as obtained from the
measurements of the polarization components. Top: the reflectivities
$r_R^{(1,2)} = |R_R^{(1,2)}|^2$. Bottom: the absolut reflection phases 
$\phi_R^{(1,2)}$ are shown assuming $\phi_R=-\pi $ at $q=0$. The solid 
curves correspond to the physical solution.

\vspace{0.5cm}
{\bf Figure 4}\\
The potential profile of the sample obtained by inversion of the 
extracted values of the reflection coefficient in the momentum range
$-1.0 \leq q \leq 1.0$\ nm$^{-1}$ shown in Fig. 3. The real and the
imaginary parts of the reconstructed profile are shown by thick solid
and dashed lines, respectively. For reference the original profile 
is shown by thin solid lines.



\newpage

\begin{center}

\begin{figure}[htb]
\centerline{\epsfig{file=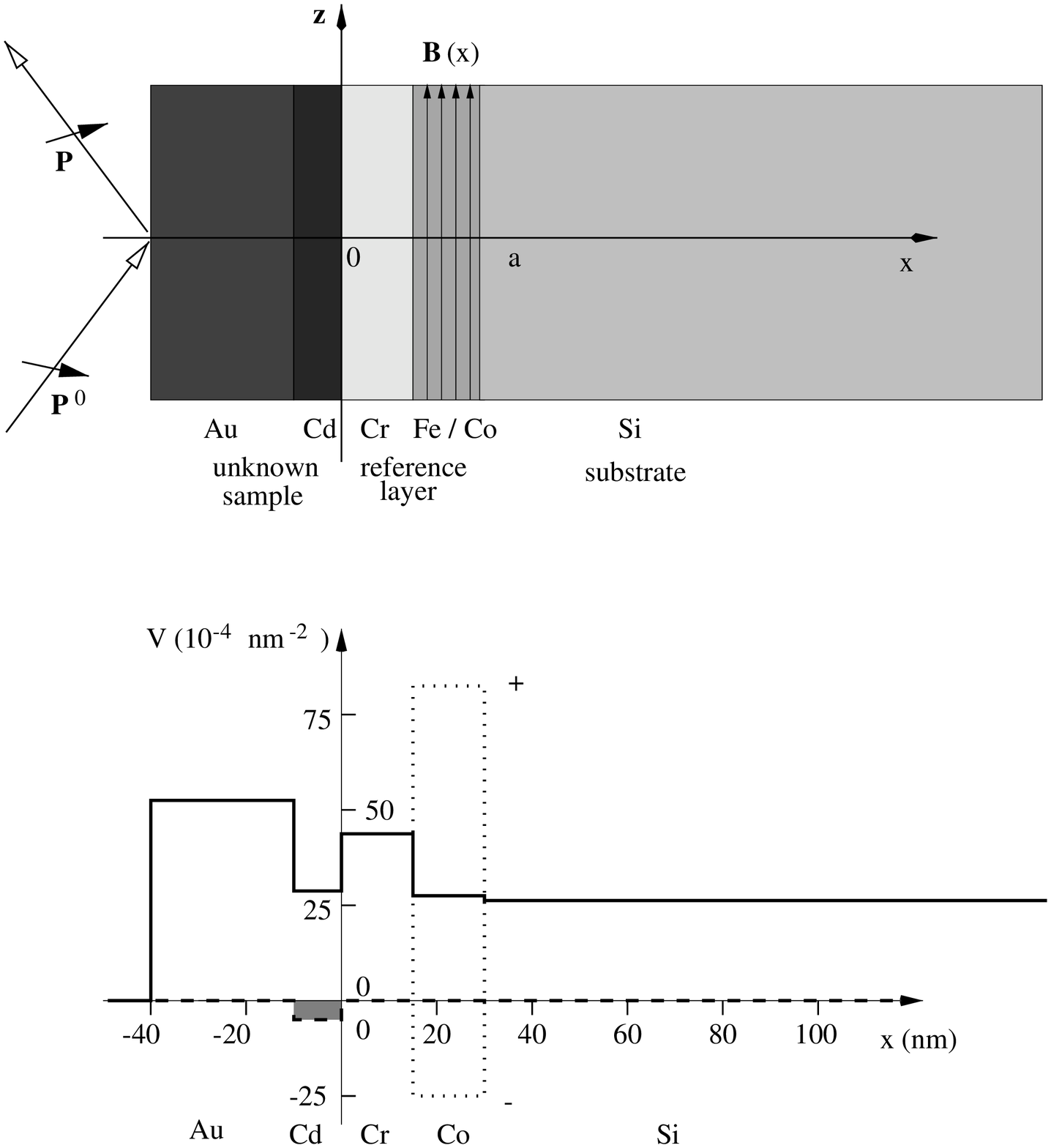,width=12.0cm}}
\end{figure}

\vspace{3cm}

{\large {\bf Figure 1}}

\newpage

\begin{figure}[htb]
\centerline{\epsfig{file=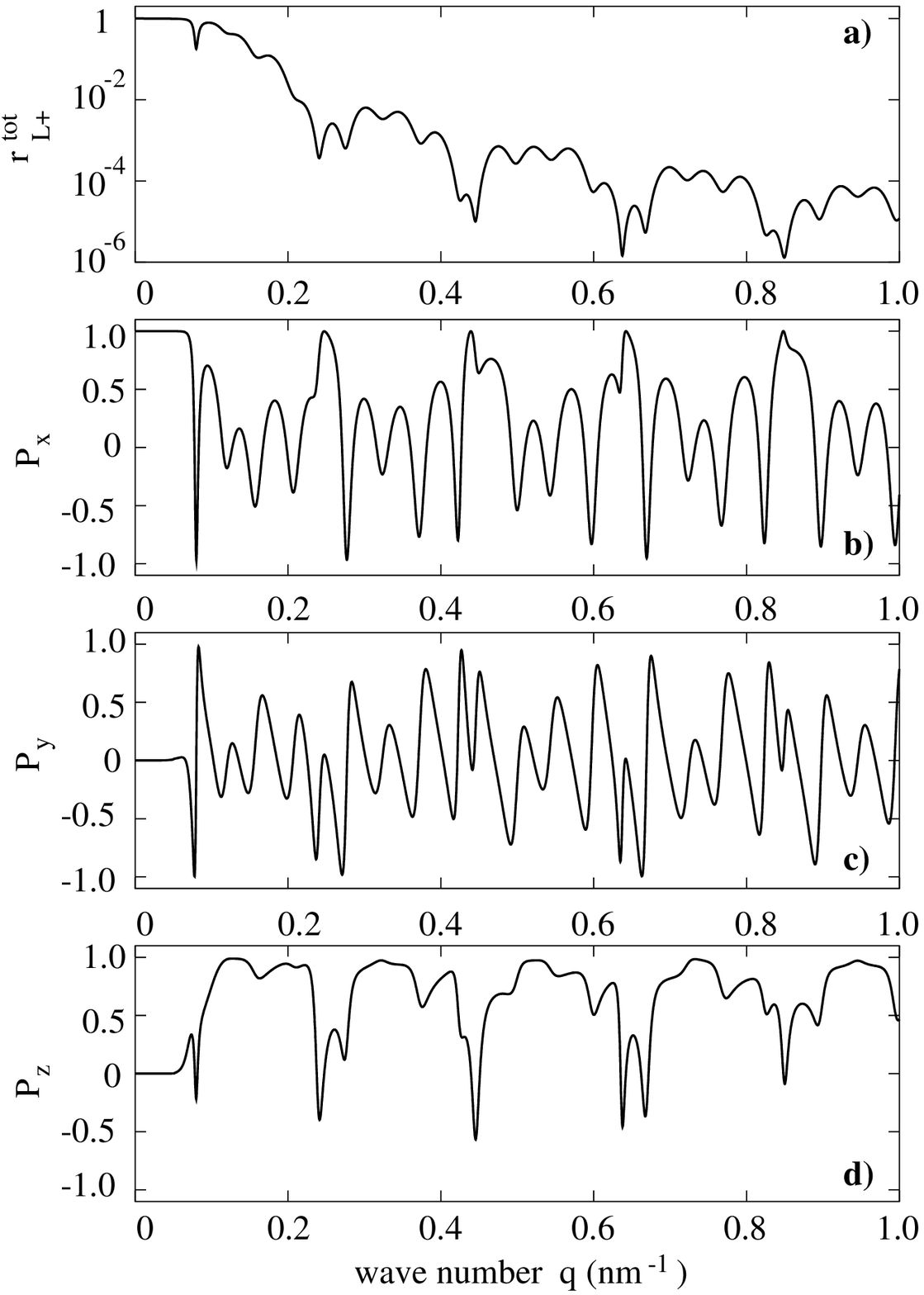,width=12.0cm}}
\end{figure}

\vspace{3cm}

{\large {\bf Figure 2}}

\newpage

\begin{figure}[htb]
\centerline{\epsfig{file=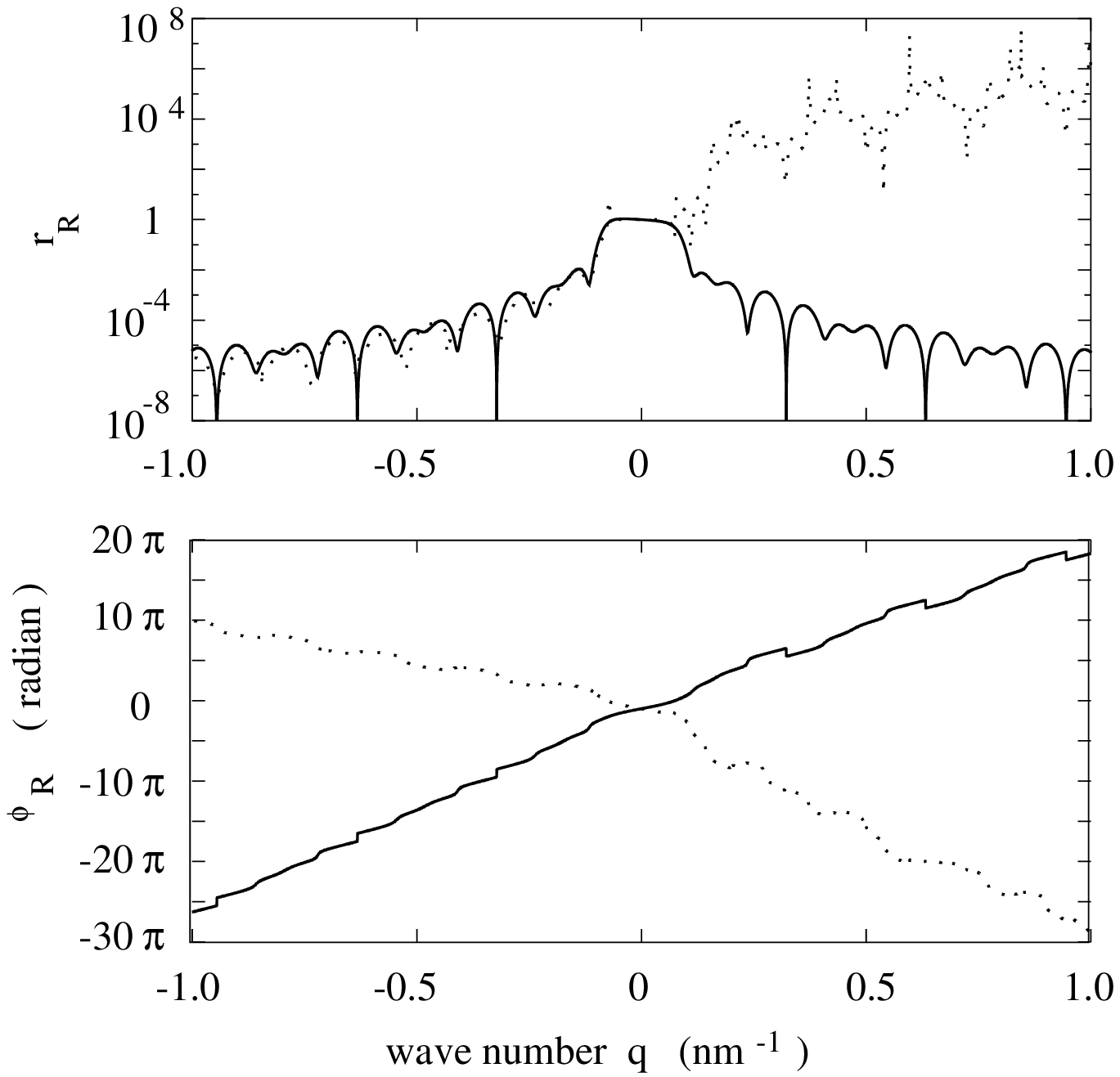,width=12.0cm}}
\end{figure}

\vspace{3cm}

{\large {\bf Figure 3}} 

\newpage

\begin{figure}[htb]
\centerline{\epsfig{file=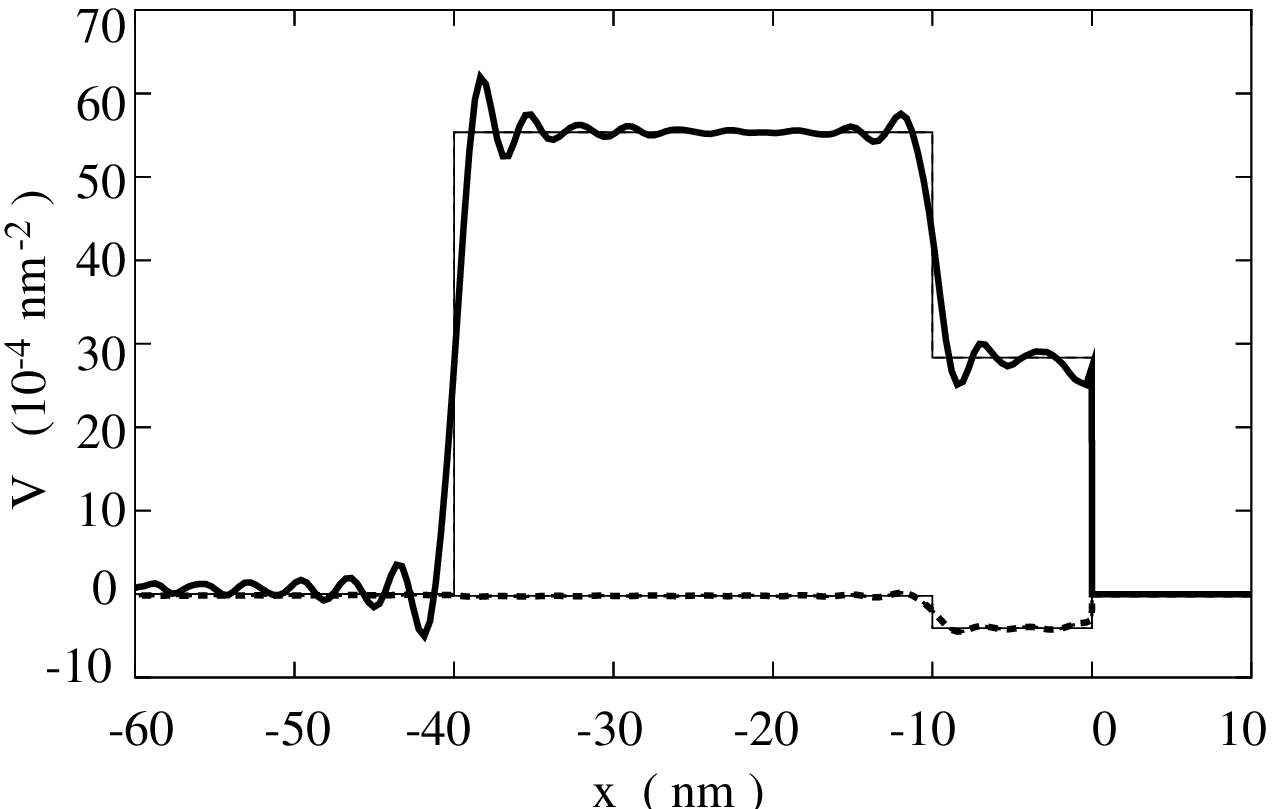,width=12.0cm}}
\end{figure}

\vspace{3cm}

{\large {\bf Figure 4}}
  
\end{center}

%

\end{document}